Stable Lithium Electrodeposition in Liquid and Nanoporous Solid Electrolytes

Yingying Lu, Zhengyuan Tu, and Lynden A. Archer*

[*]Prof. L. A. Archer

School of Chemical and Biomolecular Engineering, Cornell University, Ithaca, NY  14853-5201
Email: laa25@cornell.edu

Abstract: Rechargeable lithium, sodium, and aluminum metal-based batteries are among the most versatile platforms for high-energy, cost-effective electrochemical energy storage. Non-uniform metal deposition and dendrite formation on the negative electrode during repeated cycles of charge and discharge are major hurdles to commercialization of energy storage devices based on each of these chemistries. A long held view is that unstable electrodeposition is a consequence of inherent characteristics of these metals and their inability to form uniform electrodeposits on surfaces with inevitable defects. We report on electrodeposition of lithium in simple liquid electrolytes and in nanoporous solids infused with liquid electrolytes. We find that simple liquid electrolytes reinforced with halogenated salt blends exhibit stable long-term cycling at room temperature, often with no signs of deposition instabilities over hundreds of cycles of charge and discharge and thousands of operating hours. We rationalize these observations with the help of surface energy data for the electrolyte/lithium interface and impedance analysis of the interface during different stages of cell operation.  Our findings provide support for an important recent theoretical prediction that the surface mobility of lithium is significantly enhanced in the presence of lithium halide salts.

High energy and safe electrochemical storage are critical components in multiple emerging fields of technology where portability is a requirement for performance and large-scale deployment. From advanced robotics, autonomous aircraft, to hybrid electric vehicles, the number of technologies demanding advanced electrochemical storage solutions is rising. The rechargeable lithium ion battery (LIB) has received considerable attention because of its high operating voltages, low internal resistance and minimal memory effects[1-7]. Unfortunately LIBs are currently operating close to their theoretical performance limits due to the relatively low capacity of the anode $LiC_6$ and the lithiated cathode materials ($LiCoO_2$ and $LiFePO_4$) in widespread use. It has long been understood that a rechargeable lithium metal battery (LMB), which eschewed the use of a carbon host at the anode can lead to as much as a ten-fold improvement in anode storage capacity (from 360 mAh $g^{-1}$ to 3860 mAh $g^{-1}$) and would open up opportunities for high energy un-lithiated cathode materials such as sulfur and oxygen, among others[8-10]. Together, these advances would lead to rechargeable batteries with step-change improvements in storage capacity relative to today's state of the art LIBs.

A grand challenge in the field concerns the development of electrolytes, electrode, and battery systems configurations that prevent uneven electrodeposition of lithium and other metal anodes, and thereby eliminate dendrites at the nucleation step.[1] It is understood that without significant breakthroughs in this area, the promise of LMBs, as well as of storage platforms based on more earth abundant metals such as Na and Al metal cannot be realized. Specifically, after repeated cycles of charge and discharge growing metal dendrites can bridge the inter-electrode space, producing internal short circuits in the cell. In a volatile electrolyte, ohmic heat generated by these shorts may lead to thermal runaway and catastrophic cell failure, which places obvious safety and performance limitations on the cells. The ohmic heat generated during a short-circuit may also locally melt dendrites to create regions of "orphaned" or electrically disconnected metal that results in a steady decrease in storage capacity as a battery is cycled. Lithium-ion batteries (LiB) are designed to remove these risks by hosting the lithium in a conductive carbon host at the anode. However, the small potential difference that separates lithium insertion into versus plating onto carbon can potentially lead to similar failure modes in an overcharged or too quickly charged LiB. Thus, the need for materials that prevent non-uniform electrodeposition of metals such as Li is also implicit in new fast charging LIB technology targeted for electric-drive vehicles.

Researchers have for decades considered many approaches to stabilize lithium electrodeposition on metallic anodes[11-25]. Of these approaches, all solid-state batteries based on solid, ceramic electrolytes are considered by far the most promising, both from the perspective of their inherent safety and from theory, which indicate that a ceramic electrolyte with modulus above the shear modulus of the metallic anode can prevent dendrites from crossing the inter-electrode space. A persistent, vexing problem with ceramic electrolytes is that their room-temperature ionic conductivity rarely reach levels commonplace in liquid electrolytes and required for normal battery operation. This problem can to some extent be managed by reducing the thickness of the solid electrolyte or by operating the batteries at elevated temperature. However these changes reveal other, more serious shortcomings, which have been most clearly demonstrated in studies of high-temperature sodium metal cells[26,27]. In these cells it was found that even at temperatures where the sodium anode is a low-modulus liquid and a high-modulus, solid sodium-beta-alumina ceramic is used as the electrolyte, metal dendrites form at the electrode/electrolyte interface and ultimately proliferate through stress and corrosion-induced cracks in the ceramic. More recent studies by Tu *et al.*[14] suggest that the need for high operating temperatures can be removed by making use of nanoporous ceramics that host a low-volatility and electrochemically stable liquid electrolyte in the pores. The authors report that these electrolytes provide a combination of solid-like mechanical modulus and liquid-like bulk and interfacial conductivities at room temperature. And, when employed in LMBs, they substantially increase the lifetime of cells cycled at low and moderate current densities, but postmortem inspection of the surface of the porous ceramic host reveal that, as in the case of the bulk ceramics, dendrites are still able to nucleate and proliferate on the surface, but appear unable to penetrate through the pores of the porous material.

A longstanding puzzle in the field is that secondary batteries based on some metals (e.g. Mg) show no evidence of electrode instability and dendrite formation under deposition conditions where dendrites form and proliferate in others, such as Li[28]. At low surface deposition rates, thermodynamic and surface forces determine whether electrodeposited atoms preferentially form the low dimensionality, fiber-like structures, which lead to dendrites, or whether they form higher dimensional crystalline phases. Whereas at the intermediate and high surface deposition rates common in batteries, the mobility of atoms at the interface determines whether smooth or rough electrodeposits are created. Density functional theoretical analysis of Mg and Li electrodeposits at a vacuum metal interface reveal that Mg-Mg bonds are on average 0.18 eV stronger compared to a Li-Li[29]. This means that under the same deposition conditions, the probability of a lower dimension, fiber-like Mg deposit spontaneously transforming to a higher-dimension crystal is more than 1000 times higher than that for the corresponding transition in lithium. In electrolytes, these differences are only slightly altered by the interfacial tension, which is orders of magnitude lower, perhaps explaining why Li surfaces are more prone to nucleate dendrites irrespective of the electrolyte. A surprising and heretofore unexplored prediction from recent joint density functional theoretical (JDFT) calculations by Arias and co-workers[30,31] is that the presence of halide anions, particularly fluorides, in an electrolyte produce as much as a 0.13 eV reduction in the activation energy barrier for Li diffusion at an electrolyte-lithium metal electrode interface. If correct, this means that it should be possible to increase the surface diffusivity by more than two orders of magnitude, which may lead to large improvements in the stability of Li electrodeposition and dendrite suppression in simple liquid electrolytes.

We herein report on the stability of lithium electrodeposition in common liquid electrolytes reinforced with halogenated lithium salts. Remarkably, we find that consistent with expectations from the JDFT calculations, premature cell failure by dendrite growth and proliferation can be essentially eliminated in plate-strip type experiments even at high operating current densities. In more aggressive, high-rate polarization experiments, we find levels of dendrite suppression in room temperature liquid electrolytes that are superior to all previous reports from elevated temperature studies of polymer and other solid-state electrolytes long thought to be essential for developing reliable LMBs. Experimental characterization of the interfacial tension and impedance at the electrolyte-lithium metal interface confirm that the interfacial mobility is a strong decreasing function of halogenated lithium salt and is the most likely source of the improved stability of Li electrodeposits in liquids.

Electrolytes containing 1 M Li$^+$ cations were studied in two configurations: (i) In liquid form; and (ii) as liquids infused in nanoporous solids. Electrolytes employed in both situations were created by blending pre-determined amounts of halogenated lithium salts and lithium bis(trifluoromethanesulfonyl)imide (LiTFSI) in a low volatility propylene carbonate (PC) liquid host. To explore consequences of our observations on lifetime of lithium metal batteries, we also performed a small number of studies using blends of lithium fluoride (LiF) and lithium hexafluorophosphate (LiPF$_6$) in a 50/50 blend of ethylene carbonate (EC) and diethylene carbonate (EC:DEC). Because the most impressive enhancements in interfacial mobility predicted by JDFT are for electrolytes containing LiF, this first communication will focus on these materials. **Figure 1a** reports the DC conductivity for LiF+LiTFSI/PC as a function of LiF mole fraction in the electrolytes. It is apparent that at low LiF concentrations, DC conductivities close to the measured values for a LiTFSI/PC liquid electrolyte control are found. At LiF concentrations above 3 mol percent, the conductivity falls with increasing LiF content and the shape of the conductivity-versus-temperature profiles are seen to become flatter, but for all compositions studied, room-temperature conductivity well above $10^{-3}$ S cm$^{-1}$ are observed. A lower bulk electrolyte ionic conductivity upon addition of LiF is consistent with expectations based on the reduced dissociation of the salt, relative to LiTFSI, and consequent lower population of mobile ions in solution. The inset to the figure shows the effect of LiF on the wettability/contact angle (right axis) and surface energy (left axis) of the electrolyte with a lithium metal surface (see supplemental information **Figure S1** & **Table S1**). The measurements were performed using a home-built contact angle goniometer enclosed in an argon-filled chamber. It is apparent from the figure that addition of LiF causes a non-monotonic decrease in contact angle and a commensurate rise in interfacial energy. Later, we will show that electrodeposition of lithium metal in these electrolytes produce isolated mushroom-like structures of diameter around 40 μm. The increase in surface energy produced upon addition of LiF to the electrolytes are therefore many orders of magnitude lower than the differences in bonding energy between Mg-Mg and Li-Li atoms to significantly change the tendency of Li to form lower dimensional dendritic structures.

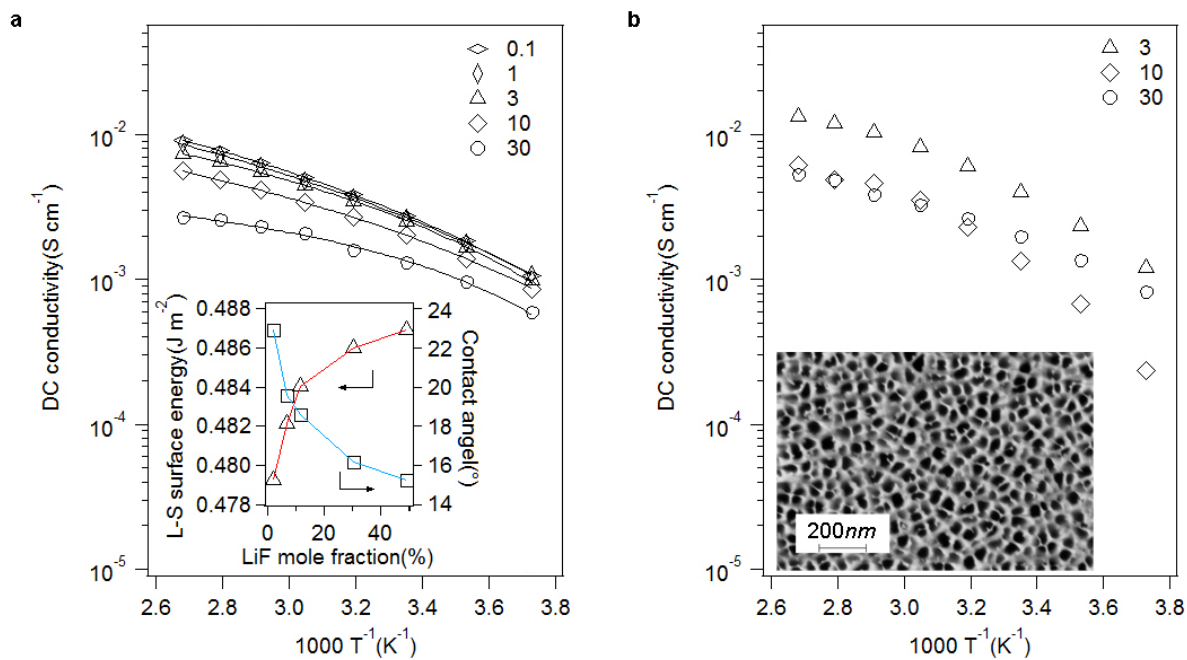

**Figure 1| DC ionic conductivity of LiF+LiTFSI/PC with various LiF mole fractions as a function of temperature. a**, Without alumina/PVDF membrane. The solid lines are Vogel-Fulcher-Tammann (VFT) fits for the temperature-dependent ionic conductivity. The parameters from the VFT fitting are shown in Table S2. The inset shows the liquid-solid surface energy and contact angle as a function of LiF mole fraction. **b**, With alumina/PVDF membrane. The SEM image shows the nanostructure of the alumina membrane with pore diameter around 40nm.

**Figure 1b** reports the DC conductivity for nanoporous solid electrolytes created by infusing LiF+LiTFSI/PC into nanoporous $Al_2O_3$/PVDF monoliths (see lower inset) with a nominal pore diameter of 40nm. The detailed preparation protocols for these electrolytes are provided in the supplementary materials section. It is apparent from **Figure 1b** that while the effect of LiF composition on conductivity is more complex than for the liquid electrolytes, over the range of LiF compositions studied the electrolytes again exhibit room-temperature conductivities above $10^{-3}$ S cm$^{-1}$; underscoring their suitability as room-temperature electrolytes for lithium batteries. Electrochemical stability of LiF-containing electrolytes was characterized by cyclic voltammetry and the results reported in the supplementary materials section (**Figure S2**). With 30mol% LiF, the width of electrochemical stability window is observed to increase measurably. The peak in the voltammogram at around 4.1 V *vs.* Li/Li$^+$ in the first cycle is in fact consistent with formation of a passivation film on the electrode that protects the electrolyte.

We investigated electrodeposition of Li in the liquid and nanoporous LiF+LiTFSI/PC based electrolytes using galvanostatic cycling of Li|LiF+LiTFSI/PC|Li symmetric lithium cells in which the lithium

striping/plating process is cycled over three-hour charge and discharge intervals designed to mimic operation in a LMB. The cells are configured to ensure that during each three-hour period sufficient lithium is transported between electrodes to create a dendrite bridge in the inter-electrode space to short-circuit the cells. The cells also do not include a separator and, once formed, the only resistance to dendrites bridging the inter-electrode spacing is provided by the intervening liquid electrolyte. **Figure 2a** compares the voltage profiles observed in symmetric cells containing electrolytes with and without LiF at a fixed, high current density of 0.38 mA cm$^{-2}$.

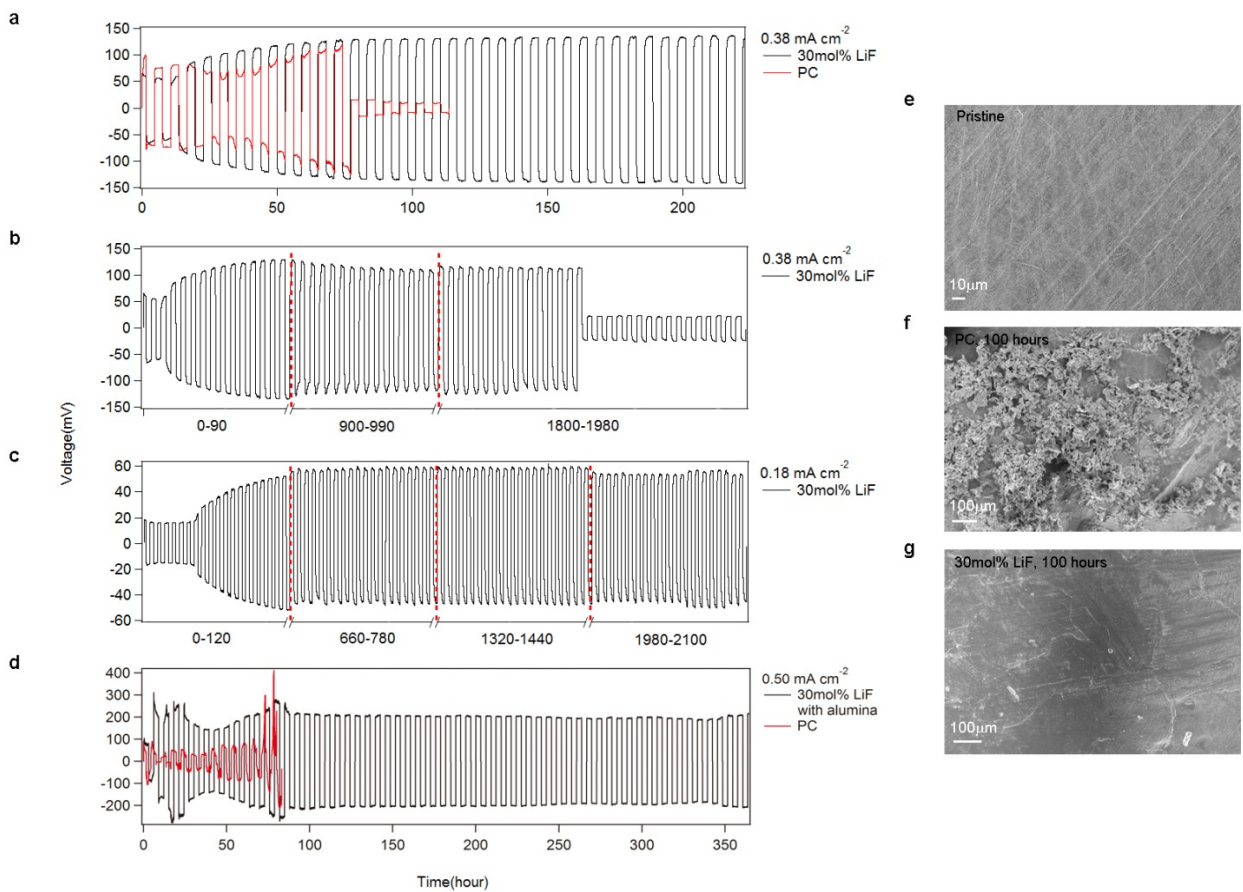

**Figure 2 | Voltage versus time for a symmetric lithium cell with each half cycle lasts 3 hours. a**, Initial voltage profiles of 30mol% LiF+LiTFSI/PC (black) and LiTFSI/PC (red) electrolytes at a current density of 0.38mA cm$^{-2}$. **b**, Voltage profile of 30mol% LiF+LiTFSI/PC electrolyte at a current density of 0.38mA cm$^{-2}$ before observing cell short-circuits. **c**, Typical voltage profile for LiF+LiTFSI/PC electrolytes at lower current densities (less than 0.2mA cm$^{-2}$). **d,** Initial voltage profiles of 30mol% LiF+LiTFSI/PC (black) and LiTFSI/PC (red) electrolytes with alumina/PVDF membrane at a current density of 0.50mA cm$^{-2}$. The initial voltage disturbance is due to the electrolyte consumption and SEI layer formation. The voltage reaches a stable plateau after 80 hours and lasts for over 350 hours. Such stable performance at high current density originate from two factors: 1) the LiF additive stabilizes the lithium deposition and forms a flat surface, which are in favor of steady battery usage; 2) the high modulus of alumina separator

prevents the dendrite proliferation and avoids the short-circuit. SEM analyses: **e**, Pristine lithium anode before galvanostatic cycling. **f**, Lithium anode in contact with LiTFSI/PC electrolyte after 100-hour cycling at 0.38mA cm$^{-2}$. **g**, Lithium anode in contact with 30mol% LiF+LiTFSI/PC electrolyte after 100-hour cycling at 0.38mA cm$^{-2}$.

The figure shows that cells that do not contain LiF in the electrolyte exhibit a large and irreversible drop in voltage consistent with catastrophic failure by a dendrite-induced short-circuit, in as little as 75 hours of operation (i.e. less than 13 cycles of charge and discharge). In contrast, cells containing 30 mol% LiF in the electrolyte cycle stably for more than 1800 hours (300 cycles of charge and discharge) before succumbing to failure in the same manner. This nearly 25-fold enhancement in cell lifetime achieved upon addition of LiF to a liquid electrolyte is considerably higher than any previous report for cells in which solid polymers[18], composites[12,13,15] and other mechanical agents are used to protect lithium metal electrodes against premature failure by dendrite-induced shorts. It is also significant that the current experiments are performed at substantially higher current densities than those reported for solid polymer or ceramic electrolytes and at room temperature. **Figure 2c** reports voltage profiles for cycling experiments performed at comparable current densities as in previous studies using polymers and other mechanical agents. Remarkably, even after 2100 hours of continuous operation, the cell shows no evidence of failure. **Figure 2d** reports a similar result for cells based on nanoporous membranes infused with liquid electrolytes, but cycled at a very high current density of 0.5 mA cm$^{-2}$. While cells with the control LiTFSI/PC electrolyte are seen to quickly fail, those containing LiF in the electrolyte settle down over a period of around 75 hours and cycle stably for more than 350 hours.

**Figures 2e-g** are scanning electron micrographs of the lithium metal electrode surface before cycling **(e)**, after 100 hours of cycling in a LiTFSI/PC control electrolyte **(f)**, and after 100 hours of cycling in a LiF+LiTFSI/PC electrolyte containing 30 mol% LiF **(g)**. It is evident from the figure that the improved lifetimes of the cells containing LiF coincides with the observation of virtually pristine Li metal electrodes after extended cycling.

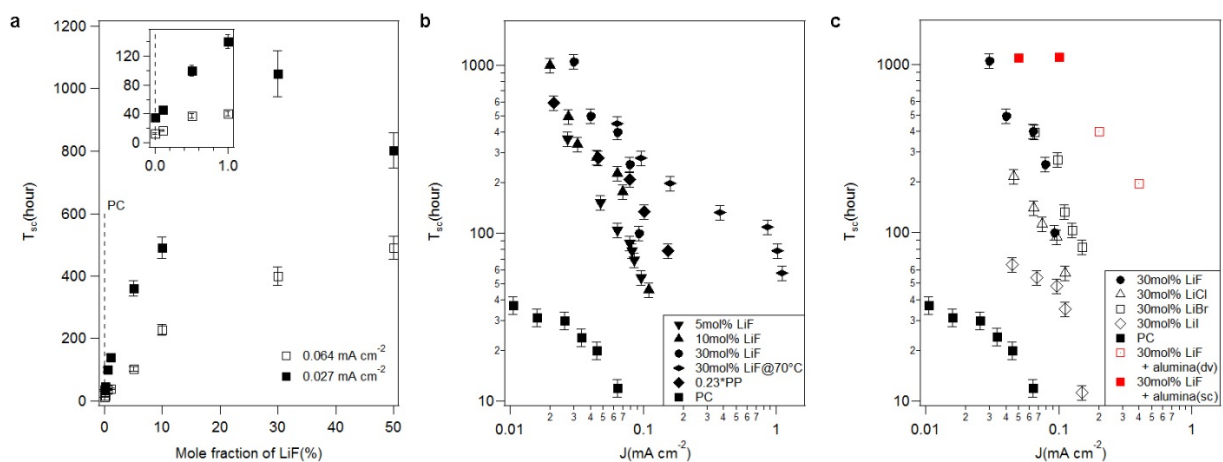

**Figure 3| Short-circuit time T$_{sc}$ from galvanostatic polarization measurements for symmetric lithium cells. a**, T$_{sc}$ as a function of LiF mole fraction at 0.027mA cm$^{-2}$, 0.064mA cm$^{-2}$. **b**, T$_{sc}$ as a function of current density *J* for various LiF concentrations and for PP-TFSI. **c**, T$_{sc}$ as a function of current density J

for different lithium halides with 30mol % of the halide. The red markers are used to represent results for cells based on nanoporous alumina/PVDF membranes infused with LiF+LiTFSI/PC electrolytes. The filled red symbols indicate the cells that short-circuit while the open red symbols represent the ones that diverge instead of short. Measurements were conducted at room temperature unless indicated.

Unidirectional galvanostatic polarization of symmetric lithium cells provides a convenient, accelerated-testing scheme for assessing the stability of lithium metal electrodes during electrodeposition. In this approach, lithium is continuously striped from one electrode and plated on the other until the cell fails by consumption of all of the lithium or as a result of a dendrite-induced short-circuit. A constant current density is applied to the cell and the corresponding voltage profile is obtained as a function of time (**Figure 4a**). The time ($T_{sc}$) at which a sharp drop-off in the potential is observed provides an estimate for its lifetime. Because there is no pause in the deposition, as occurs when the direction of the current is reversed in the cyclic plate-strip experiment discussed in the last section, there is no opportunity for defects produced by instability in one deposition cycle to heal before they nucleate dendrites that ultimately short circuit the cell. Consequently, cell failure by dendrite-induced short circuits are observed on timescales as much as one order of magnitude lower than for the plate-strip cycling measurements[12,24,25].

**Figure 3a** reports measured $T_{sc}$ values as a function of LiF concentration in the electrolyte at two current densities. Consistent with the observations reported in the previous section, the figure shows that addition of LiF to a LiTFSI/PC electrolyte produces large increases in cell lifetime. The top inset shows that addition of as little as 1mol% LiF produces more than a three-fold enhancement in cell lifetime at both low (0.027 mA cm$^{-2}$) and moderate (0.064 mA cm$^{-2}$) current densities. The figure further shows that at a higher LiF contents the relationship between $T_{sc}$ and LiF composition in the electrolyte is nonlinear. At 30 mol % LiF, it is seen that more than a 30-fold enhancement in cell lifetime is achieved at either current density, confirming the earlier observations based on cyclic plate-strip experiments. The ability of LIF salt to extend cell lifetime seems to reach its maximum level at around 30mol% LiF. For higher LiF mole fraction (50mol% LiF), there is a decrease of $T_{sc}$, which might be attributed to the low DC conductivity or low mobile ion concentration. It is also difficult to polarize the cell at relatively high current density for the same reason[23].

**Figure 3b** studies the effect of current density, $J$, and temperature on $T_{sc}$ for electrolytes containing varying concentrations of LiF, including a PC electrolyte containing 23 vol% of the ionic-liquid methy-3-propylpiperidinium (PP) TFSI known for its exceptional ability to facilitate stable electrodeposition of lithium[11,12]. It is clear from the figure that both in terms of the variation of $T_{sc}$ with $J$ and the enhancements in lifetime achieved relative to the electrolyte without additives, the LiF-based electrolytes with around 30 mol % LiF perform at least as well as those containing PP TFSI. As previously reported for electrolytes containing PP TFSI, $T_{sc}$ exhibits a power-law dependence on $J$, $T_{sc} \sim J^m$, over a wide range of current densities[13,24,25] Power law exponents $m$ obtained from the data are provided in Table S2 and show no noticeable dependence on LiF composition. It is also apparent from the figure that at 70 $^o$C electrolytes containing LiF exhibit $T_{sc}$ values with little sensitivity to $J$ over a range of current densities, allowing these electrolytes to achieve 100-fold or more enhancements in cell lifetime, relative to the control electrolyte at 25 $^o$C. **Figure 3c** nicely shows that LiF is not unique and that other halogenated lithium salts, especially LiBr, are able to significantly extend lifetime of lithium metal electrodes. **Figure 3c** further shows that $T_{sc}$ values measured using nanoporous electrolytes[14] (also see **Figure S7**) containing LiF are substantially higher than those measured in any of the other systems and are virtually independent of *J*. The two open red symbols are results for cells where no short-circuiting was observed, but in which the voltage diverged as a result of all of the lithium in the stripping electrode being plated on the other electrode without creating dendrite-induced short circuiting. It is remarkable that these cells show no evidence of short-circuiting at high current densities normally inaccessible in galvanostatic polarization experiments in symmetric Li cells. Post-mortem SEM analysis for these cells are provided as supplementary **Figures S3**.

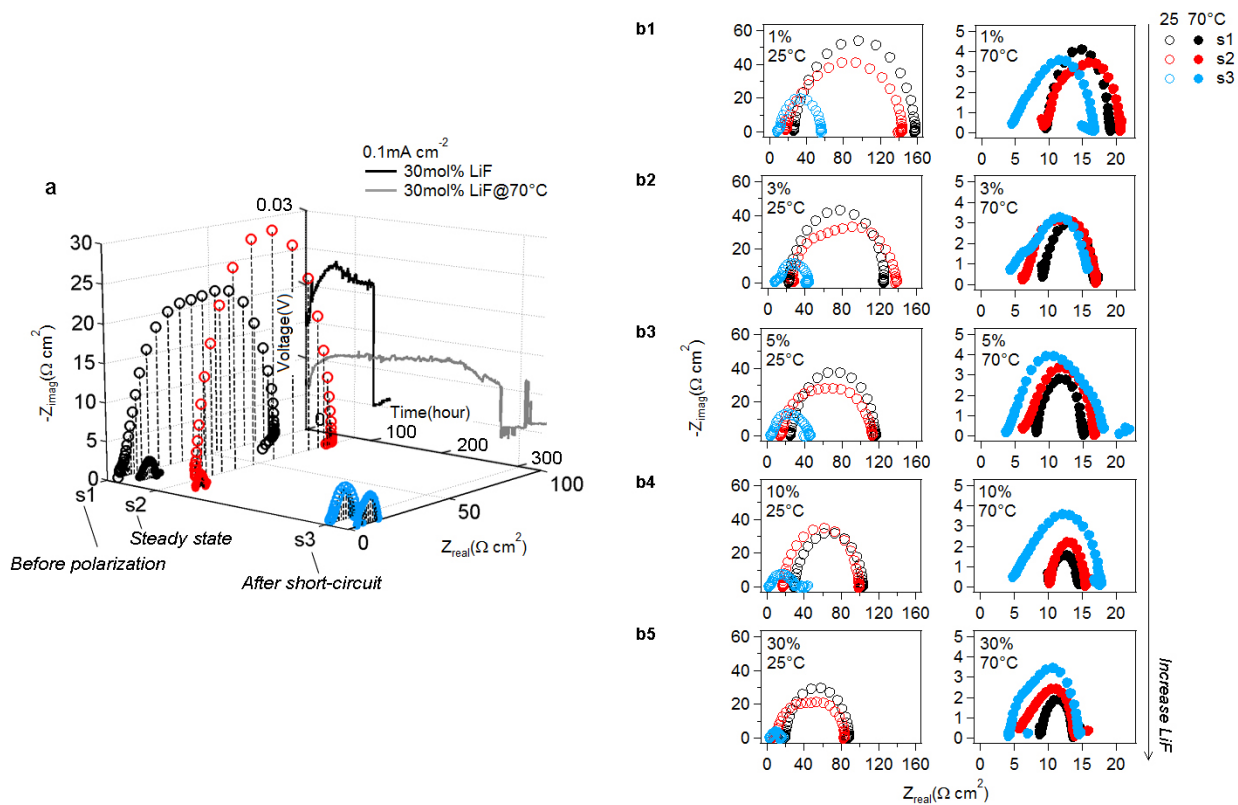

**Figure 4| Voltage profile at a fixed current density, impedance spectra of the three stages (s1: before polarization, s2: steady state, s3: after short-circuit) at 25°C and 70°C. a**, Voltage profiles and impedance spectra at 0.1mA cm$^{-2}$ for 30mol% LiF+LiTFSI/PC electrolyte. **b**, Impedance spectra for 1mol%, 3mol%, 5mol%, 10mol% and 30mol% LiF+LiTFSI/PC electrolytes. The impedance spectra with alumina/PVDF separator are reported in **Figure S8**.

Analysis of the electrode-electrolyte interface at different stages of polarization provides additional insight into the role played by LiF. Impedance spectroscopy is a frequency-domain technique that allows the complex resistance or impedance in all components of a cell (electrode, electrolyte, and their interfaces) to be determined as a function of temperature. Impedance spectra before polarization, at steady state, and after cell failure were collected and typical results reported in **Figures 4a,b**. Measurements were performed at 25°C and 70°C to characterize the effect of temperature. It is readily apparent from the figure that the interfacial impedance (related to the width of the curves) drops noticeably at the point of short-circuiting. Note that it is not possible to fit the impedance spectra by an equivalent circuit model because the surface is no longer uniform once the dendrite starts to form. **Figure 4a** compares the impedances of the three stages for 30mol% LiF+LiTFSI/PC electrolyte at 25°C and 70°C. Both the bulk (related to the lower intercept of the spectra) and interfacial impedances decrease sharply with only a 45°C temperature increase.

**Figure 4b** displays the impedance spectra for 1mol%, 3mol%, 5mol%, and 10mol% and 30mol% LiF + LiTFSI/PC electrolytes individually. At 25°C, the bulk and interfacial impedances is seen to change slightly after the onset of polarization, but as already noted drops substantially after the cell short-circuits. Electrolytes with higher LiF mole fraction have comparable bulk, but measurably lower interfacial impedances at all stages. It suggests that LiF has the ability to enhance the lithium ion diffusion primarily at the electrode/electrolyte interface. When operating at 70°C, spectra at all three stages exhibit similar bulk and interfacial impedances between 5 and 15 $\Omega$ cm$^2$ with negligible dependence of electrolyte composition. It indicates that the impedance is so small that the magnitude is almost similar to that of the short-circuited cell, which consistent with expectations based on the JDFT calculation, leads to much larger enhancements in cell lifetime. In general, the lowered impedance created by LiF leads to the extension of the cell lifetime, and the sharply reduced impedance by temperature explains the tremendous enhancement of cell lifetime at high temperature because lithium ions can easily migrate and plate on the negative electrode.

To further evaluate the suitability of LiF-added electrolytes in LMBs, more commonly used electrolytes comprised 1:1 (v:v) EC:DEC with and without LiF were investigated at room temperature using Li/Li$_4$Ti$_5$O$_{12}$ (LTO) half-cell. LTO is a no-strain material commercially utilized in electric vehicles and is capable of cycling at both low and high rates for consecutive charge and discharges[32]. In practice, even commercial LTO spinel powder yields a well-defined discharge plateau at 1.55V in carbonate electrolytes, and a discharge capacity close to the theoretical capacities (175 mAh g$^{-1}$) when accommodating lithium and negligible round-trip IR losses[33]. To characterize the effect of LiF on performance of Li/LTO half-cell, thin LTO (15 microns of active material) and thick LTO (64 microns of the active material) were studied in an accelerated procedure employing a very high current density of 2.0 mA cm$^{-2}$ (1C). For cells based on the thick LTO electrode, an activation process at 0.1C for 10 cycles was employed prior to the higher current density experiments. A two-hour charge/discharge protocol allows enough lithium to be transported during each cycle to create dendrites that are large enough to short-circuit the cells based on the thick electrode, whereas those based on the more common thin electrodes do not allow sufficient lithium transport to create a dendrite that spans the inter-electrode space.

**Figure 5(a1-b1)** show the voltage profiles obtained using the thin electrodes with and without LiF additive. Unlike the symmetric cells where the current is fixed and the voltage left unconstrained, the voltage range and current are fixed in these experiments. The onset of failure as a result of formation of dendrite shorts or regions of disconnected lithium is then expected to show up in the lifetime or capacity of the cells. It is apparent from **Figure 5(a1)** that addition of LiF to the electrolytes increases the discharge capacity, but otherwise does not alter the cycling performance of the cells. The blow-up charge and discharge curves in **Figures 5(a2)** and **Figures 5(b2)** show that the round-trip IR losses in both cells are quite minimal, as expected for LTO. The corresponding results for the thick electrodes are reported in **Figures 5c,d**. It is apparent from the voltage profiles in **Figure 5(c1)** and **Figure 5(d1)** that whereas little change in the Li-F containing electrolytes not only increases the accessible discharge capacity, but substantially improves the cycling stability of the cells. This latter feature is consistent with what one might expect from the earlier observations based on symmetric Li-Li cells which show that Li-F improves the stability of electrodeposition.

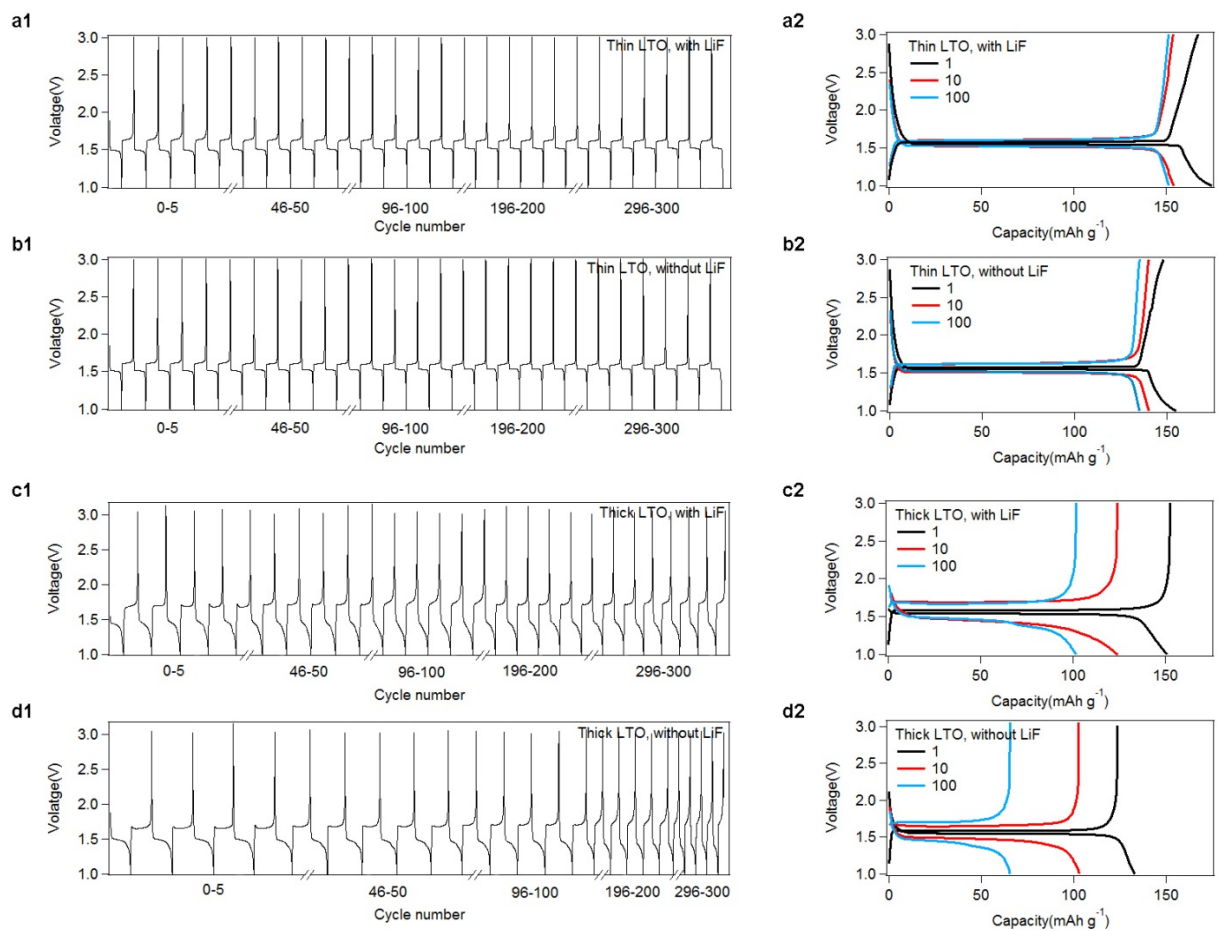

**Figure 5 | Charge-discharge characteristics of Li/Li$_4$Ti$_5$O$_{12}$ (Li/LTO) with 30mol% LiF+LiTFSI/EC:DEC and LiTFSI/EC:DEC electrolytes at room temperature.** Thin LTO electrode: Voltage *vs.* time profile for the first 5 cycles, 46-50 cycles, 96-100 cycles, 196-200 cycles and 296-300 cycles at 1C rate (0.18 mA cm$^{-2}$) with LiF (**a1**) and without LiF (**b1**). About 0.88μm lithium (charge passed=0.65C cm$^{-2}$, about 2.2μm LTO is reacted) is transported from one electrode to the other in each half cycle. Initial, 10$^{th}$, 100$^{th}$ charge-discharge profiles with LiF (**a2**) and without LiF (**b2**). Thick LTO electrode: Voltage *vs.* time profile for the first 5 cycles, 46-50 cycles, 96-100 cycles, 196-200 cycles and 296-300 cycles at 1C rate (2 mA cm$^{-2}$) with LiF (**c1**) and without LiF (**d1**). About 9.8μm lithium (charge passed=7.2C cm$^{-2}$, about 24.5μm LTO is reacted) is transported from one electrode to the other in each half cycle. Initial, 10$^{th}$, 100$^{th}$ charge-discharge profiles with LiF (**c2**) and without LiF (**d2**).

In summary, motivated by recent Joint Density Functional Theoretical calculations, which show that the presence of LiF at an electrolyte/lithium metal interface should yield large improvements in stability of Li electrodeposition, we studied physical and electrochemical properties of electrolytes containing lithium halides. Consistent with the theoretical predictions, we report for the first time that simple addition of halogenated lithium salts to a conventional low-mechanical-modulus liquid electrolyte such as PC and EC:DEC, leads to dramatic improvements in lifetime of lithium batteries utilizing metallic lithium as anode. In plate-strip symmetric cell studies, we find that Li-Li symmetric cells employing the Li halide salt reinforced electrolytes and cycled under similar conditions as reported for solid polymer and ceramic electrolytes, except at room temperature, exhibit no evidence of short circuiting.  In more aggressive polarization studies, we find that infusing the electrolytes in the pores of a nanoporous ceramic yield lithium metal electrodes that exhibit much larger lifetimes than any previously reported room-temperature battery. Our findings appear significant for at least three reasons. First, they demonstrate that the popular assumption inspired by intuition and supported by continuum modeling, that a high mechanical modulus is a requirement for an electrolyte that can stop growth and proliferation of lithium dendrites is perhaps incorrect. Second, electrolyte reinforcement by lithium halide salts provides an inexpensive easy to use strategy for stabilizing electrodeposition of lithium metal that will enable technological and scientific advances. And, third, the finding underscores the benefits of density functional and other atomistic simulation approaches for guiding materials design for batteries.

**Methods**

Pre-determined quantities of lithium halide salts and LiTFSI (LiPF$_6$) were dissolved in propylene carbonate (PC) and ethylene carbonate (EC)/diethylene carbonate (DEC) (1:1 (v:v) EC:DEC) electrolyte mixtures. The total lithium salt concentration was kept constant at 1 M. The mole fraction of LiF relative to total lithium salt was varied from 0.1-100%. To prepare the anhydrous solution, LiTFSI, LiPF$_6$, LiF, PC, EC and DEC were dried rigorously by previous reported method[12,13,34](Also see supplementary information).

Composite nanoporous alumina membranes were prepared using our previously reported approach[14]. Briefly, nanoporous alumina membranes were soaked in Polyvinylidene fluoride hexafluropropylene (PVDF-HFP)/DMF solution. Then a phase separation approach was adopted to prepare sandwich-type alumina/PVDF-HFP membranes. These membranes were further immersed in previously described LiF+LiTFSI/PC electrolyte for at least 24 hours. The coin cell configuration for alumina membrane assembly is displayed in **Scheme S1**.

The thinner Li$_4$Ti$_5$O$_{12}$ (LTO) electrodes were composed of 80% of LTO, 10% of carbon black, and 10% of PVDF binder. A pre-determined amount of N –methylpyrrolidone (NMP) was added as solvent and the resultant slurry was thoroughly mixed. Following procedure involves using a doctor blade to coat slurry on a clean copper sheet and it's rigorously dried in vacuum oven. The thicker Li$_4$Ti$_5$O$_{12}$ (LTO) electrodes (64 microns in thickness) used in the half-cell battery experiments were provided by the U.S. Department of Energy's (DOE) Cell Fabrication Facility, Argonne National Laboratory and used as received.

Symmetric lithium metal coin cells (2032 type, **Scheme S2**) were used for dielectric spectroscopy, impedance spectroscopy, cycling voltammetry, galvanostatic polarization and cycling measurements. Ionic conductivities were measured by Novocontrol N40 broadband dielectric spectrometer. The galvanostatic polarization and cycling measurements were conducted using Neware CT-3008 battery tester. Impedance spectra were measured as a function of frequency by a step heating procedure using impedance spectrometer. Cells were disassembled and the lithium metal electrodes harvested and rinsed with PC before analyzed by scanning electron microscopy (SEM, LEO1550-FESEM).

The contact angles were measured at room temperature using a Ramé-hart, Inc. Model 100-00-115 goniometer. The lithium foil was placed in a transparent environmental chamber with a rubber seal on the top. A single drop of the test liquid was placed on the substrate via a microliter syringe though the seal. The contact angle was determined six times at different positions and the average values reported.

**Acknowledgements**

This material is based on work supported as part of the Energy Materials Center at Cornell, an Energy Frontier Research Center funded by the U.S. Department of Energy, Office of Science, Office of Basic Energy Sciences under Award Number DESC0001086. This work made use of the electrochemical characterization facilities of the KAUST-CU Center for Energy and Sustainability, which is supported by the King Abdullah University of Science and Technology (KAUST) through Award# KUS-C1-018-02. Y. Lu thanks J. Jiang and Prof. C. Ober in the department of Material Science & Engineering at Cornell University for help with contact angle measurements. The thick LTO electrodes were produced at the U.S. Department of Energy's (DOE) Cell Fabrication Facility, Argonne National Laboratory. The Cell Fabrication Facility is fully supported by the DOE Vehicle Technologies Program (VTP) within the core funding of the Applied Battery Research (ABR) for Transportation Program.

**References**

1. Armand, M. & Tarascon, J.-M. Building better batteries. *Nature* **451,** 652-657 (2008).
2. Yang, P. & Tarascon, J.-M. Towards systems materials engineering. *Nat. Mater.* **11,** 560-563 (2012).
3. Xu, K. Nonaqueous liquid electrolytes for lithium-based rechargeable batteries. *Chem. Rev.* **104,** 4303-4417 (2004).
4. Kang, B. & Ceder, G. Battery materials for untrafast charging and discharging. *Nature* **458,** 190-193 (2009).
5. Morcrette, M., Rozier, P., Dupont, L., Mugnier, E., Sannier, L., Galy, J. & Tarascon, J.-M. A reversible copper extrusion-insertion electrode for rechargeable Li batteries. *Nat. Mater.* **2,** 755-761 (2003).
6. Aricò, A. S., Bruce, P., Scrosati, B., Tarascon, J.-M. & Schalkwijk, W. V. Nanostructured materials for advanced energy conversion and storage devices. *Nat. Mater.* **4,** 366-377 (2005).
7. Whittingham, M. S. Materials challenges facing electrical energy storage. *Mater. Res. Bull.* **33,** 411-419 (2008).

8. Bruce, P. G., Freunberger, S. A., Hardwick, L. J. & Tarascon, J.-M. Li-$O_2$ and Li-S batteries with high energy storage. *Nat. Mater.* **11,** 19-29 (2012).
9. Jayaprakash, N., Shen, J., Moganty, S. S., Corona, A. & Archer, L. A. Porous hollow carbon@sulfur composites for high-power lithium-sulfur batteries. *Angew. Chem. Int. Ed.* **50,** 5904-5908 (2011).
10. Tarascon, J.-M. & Armand, M. Issues and challenges facing rechargeable lithium batteries. *Nature* **414,** 359-367 (2001).
11. Bhattacharyya, R., Key, B., Chen, H., Best, A. S., Hollenkamp, A. F. & Grey, C. P. In *situ* NMR observation of the formation of metallic lithium microstructures in lithium batteries. *Nat. Mater.* **9,** 504-510 (2010).
12. Lu, Y., Korf, K., Kambe, Y., Tu, Z. & Archer, L. A. Ionic-liquid-nanoparticle hybrid electrolytes: applications in lithium metal batteries. *Angew. Chem. Int. Ed.* **53,** 488-492 (2014).
13. Lu, Y., Das, S. K., Moganty, S. S. & Archer, L. A. Ionic liquid-nanoparticle hybrid electrolytes and their application in secondary lithium-metal batteries. *Adv. Mater.* **24,** 4430-4435 (2012).
14. Tu, Z., Kambe, Y., Lu, Y. & Archer, L. A. Nanoporous polymer-ceramic composite electrolytes for lithium metal batteries. *Adv. Energy Mater.* **4,** 1300654 (2014).
15. Schaefer, J. L., Yanga, D. A. & Archer, L. A. High lithium transference number electrolytes via creation of 3-dimensional, charged, nanoporous networks from dense functionalized nanoparticle composites. *Chem. Mater.* **25,** 834-839 (2013).
16. Moganty, S. S., Srivastava, S., Lu, Y., Schaefer, J. L., Rizvi, S. A. & Archer, L. A. Ionic liquid-tethered nanoparticle suspensions: a novel class of ionogels. *Chem. Mater.* **24,** 1386-1392 (2012).
17. Moganty, S. S., Jayaprakash, N., Nugent, J. L., Shen, J. & Archer, L. A. Ionic-liquid-tethered nanoparticles: hybrid electrolytes. *Angew. Chem. Int. Ed.* **49,** 9158-9161 (2010).
18. Stone, G. M., Mullin, S. A., Teran, A. A., Hallinan Jr., D. T., Minor, A. M., Hexemer, A. & Balsara, N. P. Resolution of the modulus versus adhesion dilemma in solid polymer electrolytes for rechargeable lithium metal batteries. *J. Electrochem. Soc.* **159 (3),** A222-A227 (2012).
19. Monroe, C. & Newman, J. The impact of elastic deformation on deposition kinetics at lithium/polymer interfaces. *J. Electrochem. Soc.* **152 (2),** A396-A404 (2005).
20. Nugent, J. L., Moganty, S. S. & Archer, L. A. Nanoscale organic hybrid electrolytes. *Adv. Mater.* **22,** 3677-3680 (2010).
21. Croce, F., Appetecchi, G. B., Persi, L. & Scrosati, B. Nanocomposite polymer electrolytes for lithium batteries. *Nature* **394,** 456-458 (1998).
22. Ding, F. *et al.* Dendrite-free lithium deposition via self-healing electrostatic shield mechanism. *J. Am. Chem. Soc.* **135,** 4450-4456 (2013).
23. Chazalviel, J.-N. Electrochemical aspects of the generation of ramified metallic electrodeposits. *Phys. Rev. A* **42,** 7355-7367 (1990).
24. Rosso, M., Gobron, T., Brissot, C. Chazalviel, J.-N. & Lascaud, S. Onset of dendritic growth in lithium/polymer cells. *J. Power Sources* **97-98,** 804-806 (2001).
25. Brissot, C., Rosso, M., Chazalviel, J.-N. & Lascaud, S. Dendritic growth mechanisms in lithium/polymer cells. *J. Power Sources* **81-82,** 925-929 (1999).
26. De Jonghe, L. C., Feldman, L. & Millett, P. Some geometrical aspects of breakdown of sodium beta alumina. *Mat. Res. Bull.* **14(5),** 589-595 (1979).

27. Ansell, R. The chemical and electrochemical stability of beta-alumina. *J. Mat. Sci.* **21,** 365-379 (1986).
28. Aurbach, D. *et al.* Prototype systems for rechargeable magnesium batteries. *Nature* **407,** 724-727 (2000).
29. Ling, C., Banerjee, D. & Matsui, M. Study of the electrochemical deposition of Mg in the atomic level: Why it prefers the non-dendritic morphology. *Electrochimica Acta* **76,** 270-274 (2012).
30. Gunceler, D., Letchworth-Weaver, K., Sundararaman, R., Schwarz, K. A. & Arias, T. A. The importance of nonlinear fluid response in joint density-functional theory studies of battery systems. *Modelling Simul. Mater. Sci. Eng.* **21,** 074005 (2013).
31. Gunceler, D., Schwarz, K. A., Sundararaman, R., Letchworth-Weaver, K. & Arias, T. A. Nonlinear Solvation Models: Dendrite suppression on lithium surfaces. *16th International Workshop on Computation Physics and Materials Science: Total Energy and Force Methods*, International Centre for Theoretical Physics, Trieste (2012).
32. Brousse, T., Fragnaud, P., Marchand, R., Schleich, D. M., Bohnke, O. & West, K. All oxide solid-state lithium-ion cells. *J. Power Sources* **68,** 412-415 (1997).
33. Nakahara, K., Nakajima, R., Matsushima, T. & Majima, H. Preparation of particulate $Li_4Ti_5O_{12}$ having excellent characteristics as an electrode active material for power storage cells. *J. Power Sources* **117,** 131-136 (2003).
34. Lu, Y., Moganty, S. S., Schaefer, J. L. & Archer, L. A. Ionic liquid-nanoparticle hybrid electrolytes. *J. Mater. Chem.* **22,** 4066-4072 (2012).

Stable Lithium Electrodeposition in Liquid and Nanoporous Solid Electrolytes

Yingying Lu, Zhengyuan Tu, and Lynden A. Archer*

[*]Prof. L. A. Archer

School of Chemical and Biomolecular Engineering, Cornell University, Ithaca, NY 14853-5201 Email: laa25@cornell.edu

**Sample Preparation**

Electrolytes: A rigorous, multi-step drying protocol was used to avoid contamination of the electrolytes by residual moisture. To prepare the anhydrous solution, LiTFSI was first dried at 110 °C for 4 hours and then at 170 °C for 3 days under vacuum. Propylene carbonate (PC) was kept on 3 Å molecular sieves for at least one week and subsequently stored in an Argon-filled glovebox, MBraun Labmaster.. Ethylene carbonate (EC) was dissolved in diethyl carbonate (DEC) (1:1 v:v ratio) and kept on 3 Å sieves for at least one week and also stored in the glovebox. Pre-determined amounts of LiF+LiTFSI (LiF+LiPF$_6$) in PC (EC:DEC) was then prepared in an argon-filled glovebox.

Nanoporous alumina membranes: alumina membrane filters, Whatman Anodisc 13 with 20nm pores size, were purchased from Fisher. Polyvinylidene fluoride hexafluropropylene (PVDF-HFP, purchased from Sigma Aldrich.) was dissolved in N, N -dimethylformamide (DMF, supplied from Sigma Aldrich) at 10 wt% concentration. The viscous PVDF/DMF solution was casted on a clean glass cover slip, covered by the alumina membranes. The plate as well as the materials, were immersed in a DI water bath for 10 seconds, and a second glass cover slip was placed on top of the PVDF coated membrane to control the thickness and surface smoothness.

Supplementary Information

The resultant laminated membranes were kept in a water bath overnight before being carefully sectioned to a desired size. Composite membranes produced in this manner were transferred into a bath of dry propylene carbonate (PC), which is periodically refreshed. The membranes were further rigorously dried in 1M LiTFSI/PC solution in glovebox (MBraunn) under argon protection using a combination of lithium metal ingots and molecular sieves. The as prepared laminated alumina membranes were subsequently immersed for at least 24 hours in a 30mol% LiF+LiTFSI/PC solution containing 1M $Li^+$.

Cell Assembly: Symmetric lithium-lithium coin cell (CR2032 type) were assembled in an Argon filled glove box. To avoid any cracking of the laminated alumina membrane during the coin cell assembly, the cell configuration shown in Scheme S1 was employed. Briefly, a disc-shaped piece of lithium foil with comparable diameter as the laminated separator was placed at the bottom of the cell while a second smaller piece of lithium foil was applied to the top. This configuration is advantageous because it protects against accidental contact of the lithium electrodes during cell assembly and protects the laminated membrane from cracking when subjected to perpendicular stress during the final stages of cell assembly.

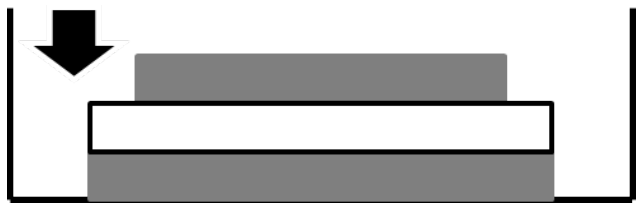

**Scheme S1| Cartoon of the specific configuration used for the Li/nanoporous membrane/Li coin cells.** Two gray plates are lithium foils while the white one designates the laminated membrane.

Supplementary Information

A more detailed configuration of the symmetric lithium cells used in to study the liquid electrolytes is provided in Scheme S2. Briefly, a Polytetrafluoroethylene (Teflon) ring is used as a separator. The outer diameter is (5/8)", inner diameter is (1/4)", and the thickness is (0.030)". The thickness of the lithium metal (Alfa Aesar) is 0.75mm. In assembling the cell, one side of the lithium metal is attached onto the Teflon ring, which creates an effective sample holder with a sealed bottom. Sufficient liquid electrolyte is slowly charged into the Teflon hole to avoid entrainment of bubbles, and the other lithium foil is carefully placed onto the other side of the separator. In this cell configuration, the liquid electrolyte in the empty space between the two electrodes provides the only barrier to lithium electrodeposits growing and proliferating in the inter-electrode space and short-circuiting the cell. Relative to lithium metal, the Teflon separator is a rigid material so the distance between the two lithium electrodes can be taken to be constant. The time for the dendrite propagation is then only related to the growth velocity.

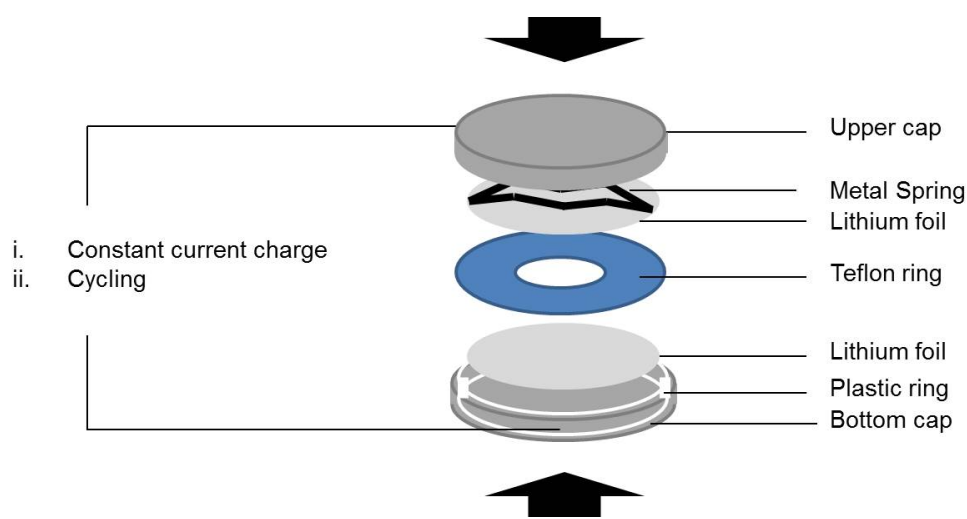

**Scheme S2| The configuration of symmetric lithium cells used in galvanostatic polarization and galvanostatic cycling measurements.**

Supplementary Information

**Sample characterization**

Polarization Experiments: Eight cells with the same electrolyte composition were prepared and the time-dependent voltage characterized at a fixed current density. At the beginning of the test, the cell was first aged for one day and rested at zero volts for 10mins. The short-circuit time was determined as the time at which the steady voltage dropped discontinuously. In all cases this drop was clean and easy to detect. Data collected for the eight cells was averaged to determine the average short-circuit time vs. current density plotted in Figure 3. The error bars in the figure are standard deviations from the average value recorded in each 8-cell sample set and are generally in the range of 8-10%. To measure the impedances at the three important stages of cell operation - before polarization, at steady state, and after cell short-circuit, cells polarized under the same conditions were terminated after 10mins rest, after the time-dependent voltage reaches steady-state and after a clear drop-off of the voltage is observed. Three cells were collected at each stage for impedance measurements. A fixed frequency range of $10^6$ to $10^{-1}$ Hz. was used for the impedance measurements.

Measuring ionic conductivity of the electrolytes: The ionic conductivity of all electrolytes was measured using frequency-domain dielectric relaxation measurements in the range $10^7$ to $10^{-1}$ Hz using a Novocontrol Broadband dielectric spectrometer. The temperature was ramped in a step profile to achieve data at -5, 10, 25, 40, 55, 70, 85 °C. The DC conductivity is extracted from the ionic conductivity vs. frequency plot at each temperature. DC ionic conductivities of LiF+LiTFSI/PC with various LiF mole fractions were measured in cells using liquid electrolytes soaked in a macroporous glass fiber separator as well as in cells using nanoporous $Al_2O_3$ infused with liquid electrolyte. Figure 1a reports the conductivity versus temperature of LiF+LiTFSI/PC soaked in glass fiber separators. The glass fiber separators were assembled in symmetric lithium coin cells for the conductivity measurement. Figure 1b shows the conductivity versus temperature of LiF+LiTFSI/PC soaked in alumina/PVDF separator. The alumina based separators were loaded between two copper plates for conductivity test.

The continuous lines through the data in Figure 1a are obtained by fitting the experimental results to the Vogel-Fulcher-Tammann (VFT) formula, $\sigma = A\exp(-B/(T-T_0))$, where $B$ is the effective activation energy barrier, in the units of absolute temperature; $T$ and $T_0$ are the measurement and reference temperatures, respectively; and $A$ is a pre-exponential factor which equals to the ionic conductivity in the high-temperature limit. The equation fits the data over the range of temperature and lithium fluoride content studied[4-6]. The short-circuit time *vs.* current density by galvanostatic polarization measurement is fitted by the following power law function[7-10]: $T_{SC} = AJ^{-m}$. The parameter $A$ is related to the diffusion coefficient of ions, mobile ion concentration, anion transport number, and ion mobility[8-10]. Values of all of these parameters are presented in Table S2.

Galvanostatic cycling experiments: For the galvanostatic cycling test, the cell was initially charged at a fixed current density for 90mins, and then discharged at the same current density for 180mins, followed with charging 180mins to continue the cycling.

Characterizing the Interfacial Energy at the electrode/electrolyte interface: The contact angle of each liquid electrolyte on the surface of clean lithium metal was measured using a goniometer in an argon-filled enclosure. Analysis of the contact angle data by Young's equation and Zisman approach[11-13] can be used to determine the surface energy as follows:

Supplementary Information

$$\gamma_L \cos\theta = \gamma_S - \gamma_{SL} \quad (1)$$

$$\cos\theta = 1 - b(\gamma_L - \gamma_C) \quad (2)$$

Where $\gamma_L$ is the experimentally determined surface energy (surface tension) of the liquid (electrolyte), $\theta$ is the contact angle, $\gamma_S$ is the surface energy of the solid (lithium: 0.52 J m$^{-2}$), $\gamma_{SL}$ is the solid/liquid interfacial energy, $b$ is the slope of the regression line and $\gamma_C$ is the critical surface tension when $\cos\theta = 1$. The two parameters $b$ and $\gamma_C$ are calculated from (2) using $\gamma_L$ values from the literature and the measured $\theta$ for pure PC and DMC. The variation of $\gamma_C$ in electrolytes with different concentrations of LiF was obtained from the regression line (Figure S1). $\gamma_{SL}$ were then calculated from (1). Note that the deviations of the measurements and Zisman calculation may apply but the results in terms of the L-S surface energy change very small due to the large lithium surface energy compared with that of a liquid electrolyte. Results from this method are displayed in Table S1.

**Table S1| Contact angles, liquid electrolyte surface tensions ($\gamma_L$), lithium surface energy ($\gamma_S$) and solid/liquid interfacial energies ($\gamma_{SL}$) of various electrolyte compositions.**

| Sample | $\theta(°)$ | $\cos(\theta)$ | Electrolyte surface tension(N m$^{-1}$) | Liquid-solid surface energy(J m$^{-2}$) |
|---|---|---|---|---|
| Pure PC | 23.0 | 0.92 | 0.0448 | 0.479 |
| Pure DMC | 7.3 | 0.99 | 0.0291 | 0.491 |
| PC/1M LiTFSI | 22.0 | 0.93 | 0.0426 | 0.480 |
| 5 mol% LiF | 19.5 | 0.94 | 0.0403 | 0.482 |
| 10 mol% LiF | 18.5 | 0.95 | 0.0381 | 0.484 |
| 30 mol% LiF | 16.0 | 0.96 | 0.0358 | 0.486 |
| 50 mol% LiF | 15.0 | 0.97 | 0.0336 | 0.487 |
| Li: 0.52 J m$^{-2}$ [1]; PC: 0.045 N m$^{-1}$ [2]; DMC: 0.0286 N m$^{-1}$ [3] | | | | |

Supplementary Information

**Table S2 | VFT fitting parameters and scaling exponent by power law fitting.** The fittings of these two equations were conducted by Origin 8.0.

| Sample | VFT: $\sigma = A\exp(-B/(T-T_0))$ | | | $T_{SC} = AJ^{-m}$ |
|---|---|---|---|---|
| | $A(S/cm)$ | $B(K)$ | $T_0(K)$ | $m$ |
| 0.1 mol% LiF | 0.081±0.0029 | 461±12 | 162±2.1 | - |
| 1 mol% LiF | 0.079±0.0011 | 485±5.1 | 156±0.88 | - |
| 3 mol% LiF | 0.035±0.0028 | 291±22 | 188±4.9 | - |
| 5 mol% LiF | - | - | - | 1.42 |
| 10 mol% LiF | 0.0033±0.0085 | 365±8.1 | 168±1.7 | 1.53 |
| 30 mol% LiF | 0.0061±0.00097 | 127±32 | 215±12 | 1.70 |
| 100 mol% LiF | 0.00052±0.000030 | 506±21 | 138±3.8 | |
| 30 mol% LiF @ 70°C | - | - | - | 0.60 |
| 30 mol% LiCl | - | - | - | 1.36 |
| 30 mol% LiBr | - | - | - | 2.10 |
| 30 mol% LiI | - | - | - | 1.27 |

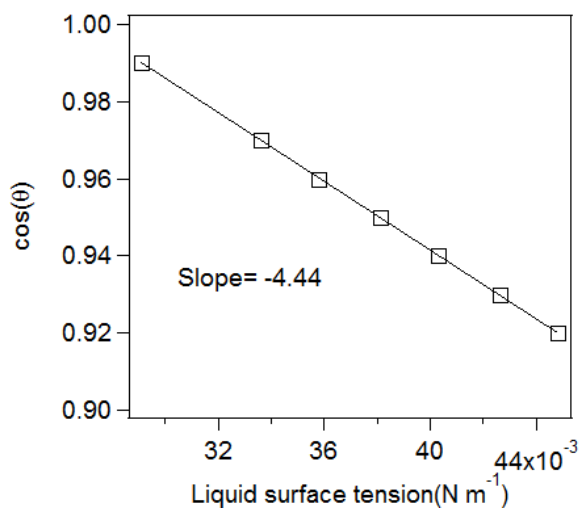

**Figure S1 | Zisman's plot ($\cos\theta$ vs. $\gamma_L$) for LiF+LiTFSI/PC, PC, DMC on lithium metal surface.** Based on Zisman approach, the contact angle of different liquids on the same surface is linearly dependent on the liquid surface tension.

Supplementary Information

Analysis of the surface tension results suggests that although the contact angle decreased by LiF concentration increment, the surface energy at the electrode/electrolyte interface is only modestly changed over the range of LiF concentrations studied. However, the general trend where a higher surface energy correlating with more stable electrodeposition is perhaps intuitive and therefore of potential interest.

There are two possible mechanisms that might explain a dependence of electrodeposition stability on the surface energy at the electrolyte/lithium interface. First, during polarization, the striped lithium ions deposit on the negative lithium electrode and create a newly formed lithium surface. For the initially deposited lithium ions, they tend to create a homogenous dispersity since the lithium surface is pristine. thus the current density is uniform on the surface. For a liquid electrolyte with enhanced wettability, the lithium ions near the negative electrode are delivered to the interface in a more uniform distribution. This allows the second, third, and subsequently deposited layers to faithfully track the topology of the first deposited Li layer, which favors a uniform deposition. In addition, because of the relatively homogenous surface morphology, the centralization of the current density along the surface is reduced, which also retards the propagation of the dendrite growth. Consequently, the cell lifetime is prolonged when the contact angle is decreased. Second, during the dendrite proliferation process, the lithium ions are more likely to deposit on dendrite tips when the surface energy is low. This is due to the energy for creating a new surface on the tip is higher than that on a flat surface. So the propagation of the dendrite would be suppressed by increasing the S-L surface energy[14-17]. The experimental results show that by lowering the contact angle or increasing the S-L surface energy, the surface roughness tends to decay.

Supplementary Information

Electrochemical Stability of the Electrolyte: Figure S2 shows the effect of LiF on the electrochemical stability of a PC-based electrolyte. The measurements were performed using a linear sweep voltammetry technique at a fixed scan rate of 0.5 mV/s, The electrochemical stability window of the electrolyte is seen from the figure to increase from 0.8V to 1.2V *vs.* Li/Li$^+$. The current peak observed at high potentials in the first cycle, but which disappeares afterwards, is consistent with the idea that a passivation film is created on the Li surface at around 4.1 V *vs.* Li/Li$^+$. In this regard, the combination of carbonate electrolytes with fluorinated lithium salt could have similar effect as fluorinated carbonate compound, which has been reported to exhibit low melting point, increased oxidation stability and less flammability[18].

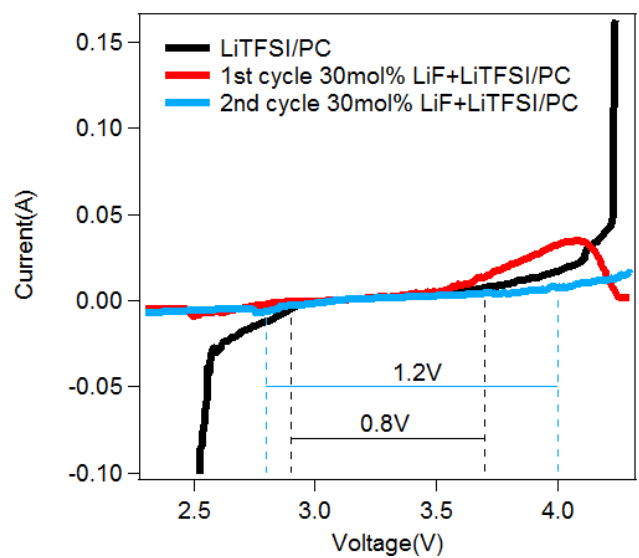

**Figure S2| Electrochemical stability window from cyclic voltammetry measurements for 1M LiTFSI/PC and 30mol% LiF+LiTFSI/PC at a rate of 0.5 mV s$^{-1}$.** The measurements were conducted in symmetric lithium cells.

Supplementary Information

*SEM Characterization of the separator and anode surfaces*: Figure S3 reports SEM micrographs of the lithium surface retrieved from the lithium/nanoporous alumina/lithium cell cycled under the most extreme current density studied 0.5 mA cm$^{-2}$ after 400 hours lithium plating/striping.

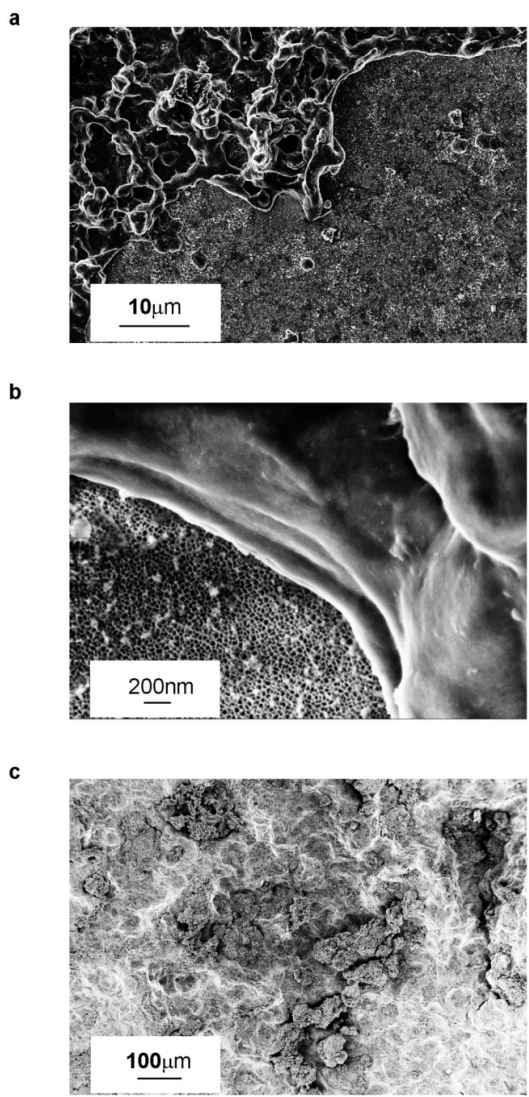

**Figure S3| SEM analysis of alumina/PVDF separator and lithium substrate after 400 hours lithium plating/striping test at 0.5 mA cm$^{-2}$. a**, Alumina/PVDF separator with lithium metal on one side. **b**, Zoom in picture of **a**. **c**, Morphology of lithium surface.

Supplementary Information

The alumina surface is sparsely covered by ramified deposits, presumed caused by dendritic electrodeposition. Considering that the pore sizes of the separator is at most 40 nm (see Figure S4), it is clear that these deposits cannot penetrate through the pores of the separator. This observation is an important confirmation of our design concept as it implies that although the pore structures in alumina allow ions to move freely, as required for high room temperature conductivity, the pores are small enough to prevent ready access and transport by metal dendrites. The morphology of lithium surface indicates that the size of lithium dendrite is much larger than the pore size, meaning that the dendrite cannot penetrate through the porous media.

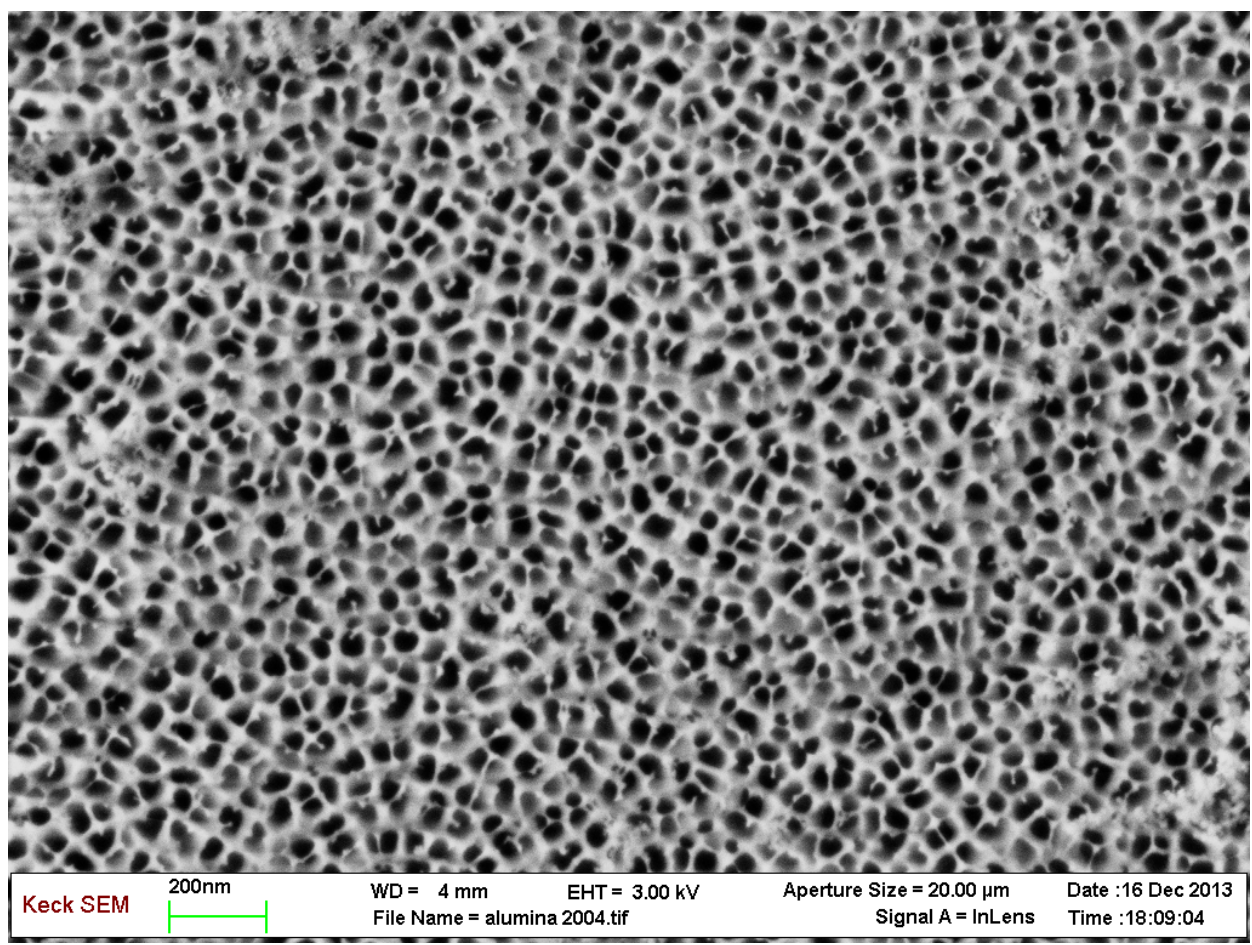

**Figure S4| SEM analysis for nanoporous alumina membranes.** The pore size of the alumina film is around 40nm.

Supplementary Information

Figure S5 are the post-mortem SEM images of the lithium negative electrodes at various polarization current densities. It is apparent that the majority of the electrodes remain quite flat (Figure S6) even after the voltage discontinuity evidences short-circuiting. At low current densities (i.e. 0.027 mA cm$^{-2}$, 0.033 mA cm$^{-2}$) small patches of mushroom-like structures are observed. These give way to thicker, needle-like structures at high current densities (i.e. 0.060 mA cm$^{-2}$, 0.082 mA cm$^{-2}$). Note that at current density as high as 0.082 mA cm$^{-2}$, it is difficult to focus all the dendrite tips in one x-y plane because the structures are thick and overlapped. The average cross section diameter of each dendrite structure arises from 7.5 to 40 microns as reducing the polarization current density. To obtain the average diameter, the smallest tips in each figure are chosen for calculation because they are the most possible dendrites to cause a short circuit. Specifically, the current density is higher on a tip with larger curvature, thus the dendrite growth rate increases.

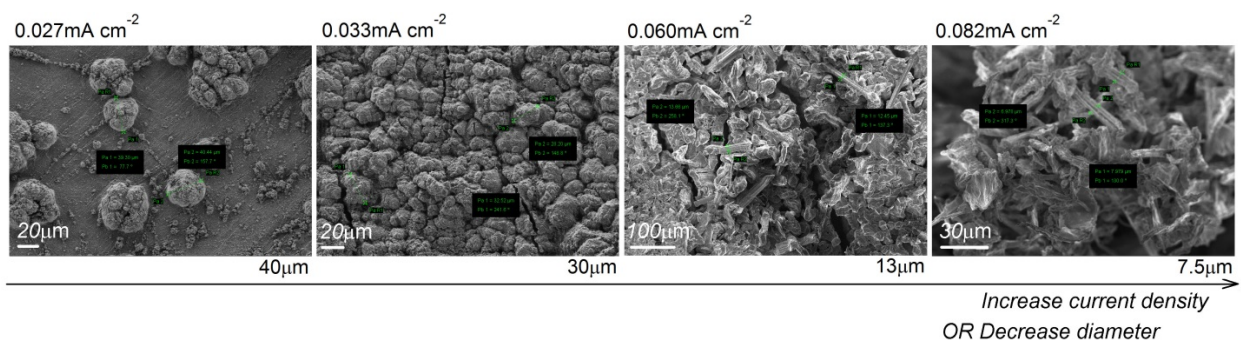

**Figure S5| Post-mortem SEM images of the lithium negative electrodes at various polarization current densities.** The average cross section diameters are shown on the bottom right of each picture.

Figure S6 shows that the majority of the lithium electrode remains quite flat after cell short-circuits in the polarization experiments even for electrodes after higher current density charging.

Supplementary Information

This means that the short circuits are likely the result of isolated dendrites. The dendrite tip is smaller under higher current density.

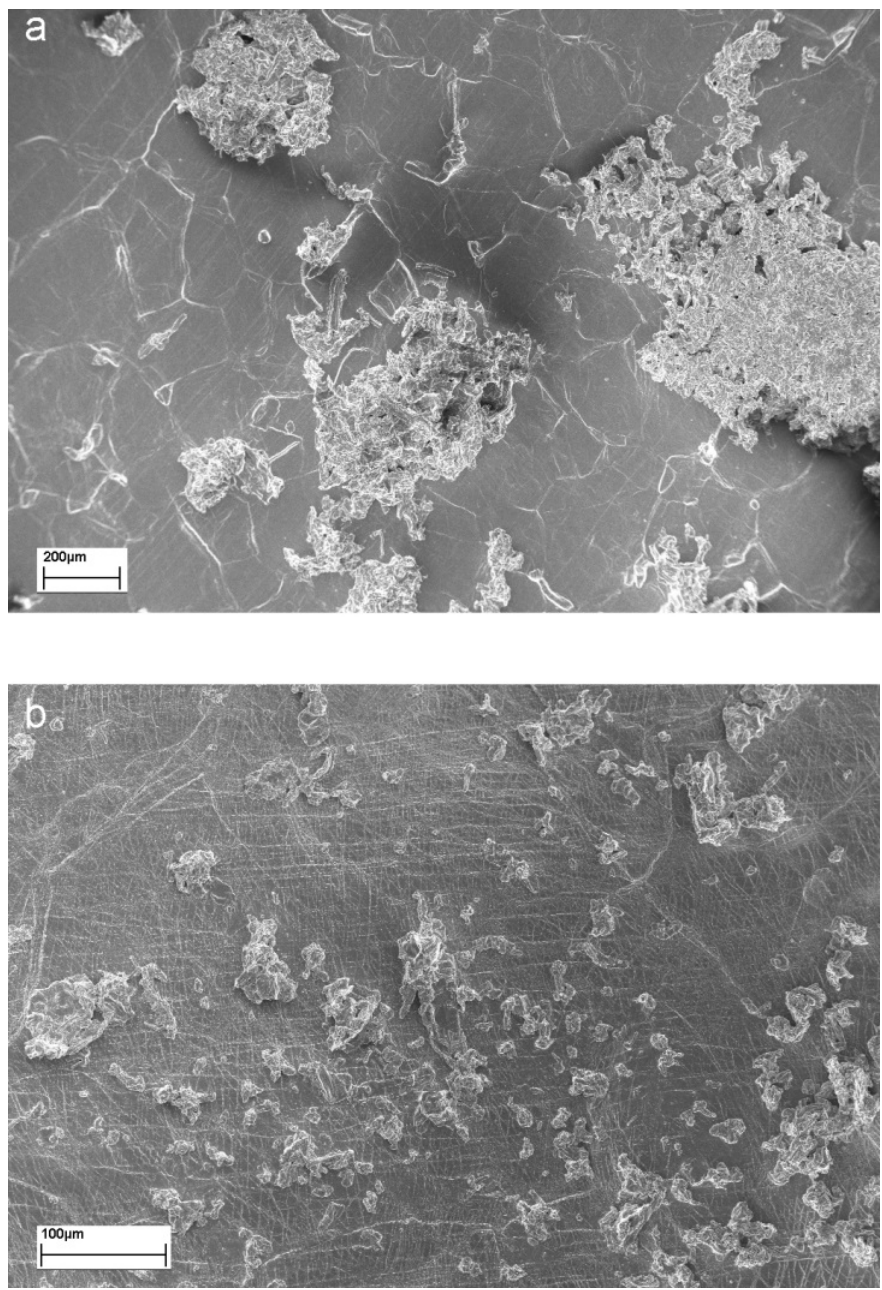

**Figure S6 | Post-mortem SEM images of the lithium negative electrodes after galvanostatic polarization measurement.** a. at fixed current density of 0.082mA cm$^{-2}$. b. at fixed current density of 0.06mA cm$^{-2}$.

Analysis of Voltage profiles during polarization of Li/nanoporous $Al_2O_3$/Li cells: As discussed in the manuscript, lithium-lithium symmetric cells utilizing a nanoporous Al2O3 separator infused with LiF+LiTFSI/PC liquid electrolyte revealed unprecedented, long cell lifetimes even at high current densities. Figure S7. are typical voltage profiles recorded for these cells. The divergent voltage evident in 5(b) imply that the short-circuit time is longer than 400 hours, the time at which the experiments were terminated.

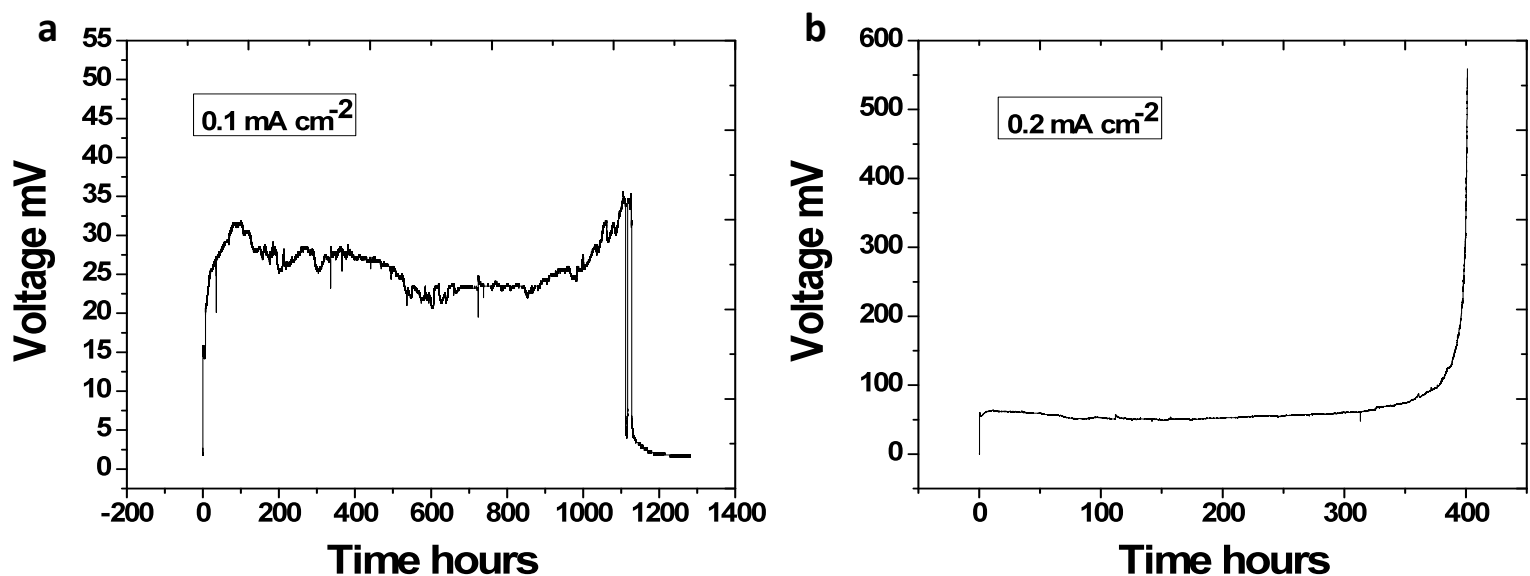

**Figure S7| Potential profile for symmetric lithium cell with 30mol% LiF+LiTFSI/PC using nanoporous alumina separator. a**, potential profile at 0.1mA $cm^{-2}$. **b**, potential profile at 0.2mA $cm^{-2}$.

Figure S8 reports the impedance of Li/30% LiF+LiTFSI/PC/Li cells as a function of temperature. As expected, both real and imaginary impedance decrease within the increase of temperature from -5 ℃ to 100 ℃. An R(RQ)W circuit model was used to extrapolate bulk and interfacial resistance. The results are displayed in Figure S8 (c). It's clear that the alumina/PVDF separator with 30% LiF additive shows quite low bulk and interfacial resistances at room temperature,

Supplementary Information

which is advantageous for battery function and stable lithium electrodeposition. The interfacial resistance unexpectedly increases at 75 ℃ and then decreases again, indicating that diffusion at the interface becomes the dominating factor that influences the charge transport.

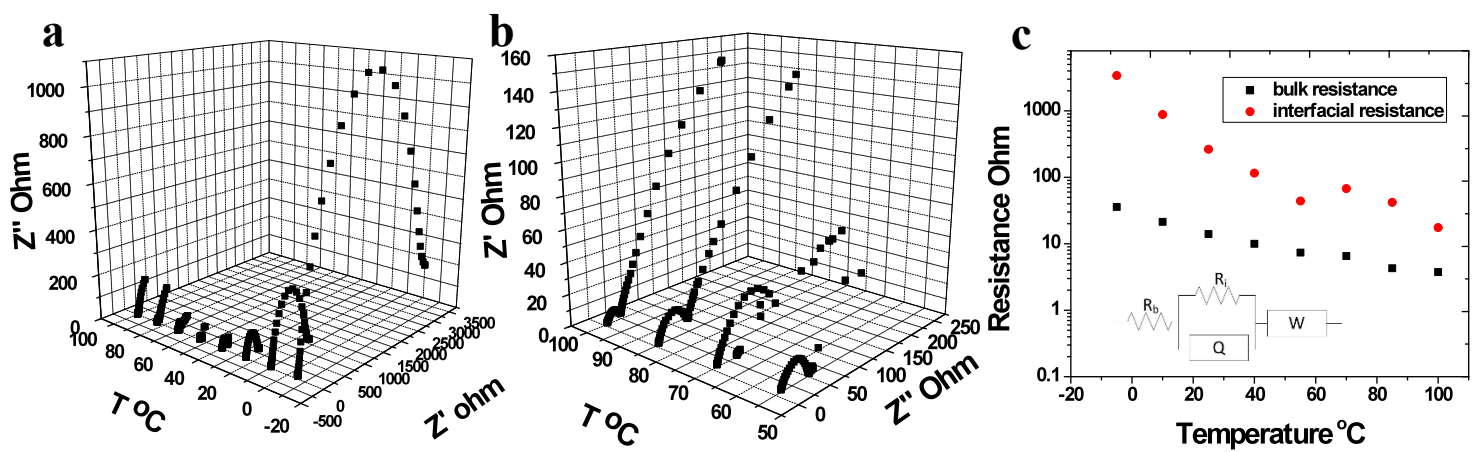

**Figure S8| Impedance spectra of 30% LiF + LiTFSI/PC in alumina/PVDF separator. a**, Impedance spectrum versus temperature of 30% LiF + LiTFSI/PC in alumina/PVDF separator. **b**, A zoom-in impedance spectrum in **a** from 55 ℃ to 100 ℃. **c**, Extrapolated bulk and interfacial resistance from **a**; the fitting circuit is shown as inset.

**References Cited**

1. Kokko, K., Salo, P. T., Laihia, R. & Mansikka, K. First-principles calculations for work function and surface energy of thin lithium films. *Surface Science* **348,** 168-174 (1996).
2. Dahbi, M., Violleau, D., Ghamouss, F., Jacquemin, J., Tran-Van, F., Lemordant, D. & Anouti, M. Interfacial properties of LiTFSI and $LiPF_6$-based electrolytes in binary and ternary mixtures of alkylcarbonates on graphite electrodes and calgard separator. *Ind. Eng. Chem. Res.* **51,** 5240-5245 (2012).

3. Deng, J., Yang, Y., He, Y., Ouyang, G. & Huang, Z. Densities and surface tensions of trimethylbenzene + dimethyl carbonate or + diethyl carbonate at 298.15 K and 313.15K. *J. Chem. Eng. Data* **51,** 1464-1468 (2006).
4. Lu, Y., Moganty, S. S., Schaefer, J. L. & Archer, L. A. Ionic liquid-nanoparticle hybrid electrolytes. *J. Mater. Chem.* **22,** 4066-4072 (2012).
5. D. I. Bower, *An Introduction to Polymer Physics,* 2002, Ch. 7.
6. Nugent, J. L., Moganty, S. S. & Archer, L. A. Nanoscale organic hybrid electrolytes. *Adv. Mater.* **22,** 3677-3680 (2010).
7. Lu, Y., Korf, K., Kambe, Y., Tu, Z. & Archer, L. A. Ionic-liquid-nanoparticle hybrid electrolytes: applications in lithium metal batteries. *Angew. Chem. Int. Ed.* **52,** 1-6 (2013).
8. Chazalviel, J.-N. Electrochemical aspects of the generation of ramified metallic electrodeposits. *Phys. Rev. A* **42,** 7355-7367 (1990).
9. Rosso, M., Gobron, T., Brissot, C. Chazalviel, J.-N. & Lascaud, S. Onset of dendritic growth in lithium/polymer cells. *J. Power Sources* **97-98,** 804-806 (2001).
10. Brissot, C., Rosso, M., Chazalviel, J.-N. & Lascaud, S. Dendritic growth mechanisms in lithium/polymer cells. *J. Power Sources* **81-82,** 925-929 (1999).
11. Zisman, W. A. Influence of constitution on adhesion. *Ind. Eng. Chem.* **55,** 19-38 (1963).
12. Gindl, M., Sinn, G., Gindl, W., Reiterer, A. & Tschegg, S. A comparison of different methods to calculate the surface free energy of wood using contact angle measurements. *Colloids and Surfaces A: Physicochem. Eng. Aspects* **181,** 279-287 (2001).
13. Dahbi, M., Violleau, D., Ghamouss, F., Jacquemin, J., Tran-Van, F., Lemordant, D. & Anouti, M. Interfacial properties of LiTFSI and $LiPF_6$-based electrolytes in binary and ternary mixtures of alkylcarbonates on graphite electrodes and calgard separator. *Ind. Eng. Chem. Res.* **51,** 5240-5245 (2012).
14. Aurbach, D. *et al.* Prototype systems for rechargeable magnesium batteries. *Nature* **407,** 724-727 (2000).
15. Ling, C., Banerjee, D. & Matsui, M. Study of the electrochemical deposition of Mg in the atomic level: Why it prefers the non-dendritic morphology. *Electrochimica Acta* **76,** 270-274 (2012).
16. Gunceler, D., Letchworth-Weaver, K., Sundararaman, R., Schwarz, K. A. & Arias, T. A. The importance of nonlinear fluid response in joint density-functional theory studies of battery systems. *Modelling Simul. Mater. Sci. Eng.* **21,** 074005 (2013).
17. Gunceler, D., Schwarz, K. A., Sundararaman, R., Letchworth-Weaver, K. & Arias, T. A. Nonlinear Solvation Models: Dendrite suppression on lithium surfaces. *16th International Workshop on Computation Physics and Materials Science: Total Energy and Force Methods*, International Centre for Theoretical Physics, Trieste (2012).
18. Zhang, Z., Hu, L., Wu, H., Weng, W., Koh, M., Redfern, P.C., Curtis, L. A. & Amine, K. Fluorinated electrolytes for 5V lithium-ion battery chemistry. *Energy. Environ. Sci.* **6,** 1806-1810 (2013).